\documentclass[a4paper,english]{article}
\usepackage{times}
\usepackage[T1]{fontenc}
\usepackage{babel}
\usepackage{color}
\usepackage{graphics}
\usepackage{setspace}
\usepackage{longtable}

\makeatletter

\providecommand{\LyX}{L\kern-.1667em\lower.25em\hbox{Y}\kern-.125emX\@}
\let\SF@@footnote\footnote
\def\footnote{\ifx\protect\@typeset@protect
    \expandafter\SF@@footnote
  \else
    \expandafter\SF@gobble@opt
  \fi
}
\expandafter\def\csname SF@gobble@opt \endcsname{\@ifnextchar[
  \SF@gobble@twobracket
  \@gobble
}
\edef\SF@gobble@opt{\noexpand\protect
  \expandafter\noexpand\csname SF@gobble@opt \endcsname}
\def\SF@gobble@twobracket[#1]#2{}

\makeatother
\begin{document}
\textcolor{black}{\begin{tabular}{c}
\\
\\
\\
\\
\textcolor{black}{Low noise temperature control: application to an
active cavity radiometer}\\
\end{tabular}}

\begin{tabular}{l}
\\
Bruno Guillet\( ^{*} \), Didier Robbes and Laurence Méchin\\
GREYC (CNRS UMR 6072)\\
Institut des Sciences de la Matière et du Rayonnement,Université de
Caen,\\
6 Bd Maréchal Juin, 14050 Caen Cedex, France\\
\( ^{*} \) author to whom correspondence should be addressed,\\
electronic mail: bguillet@greyc.ismra.fr\\
(Received\\
\end{tabular}

\begin{abstract}
\textcolor{black}{We have designed low noise temperature sensing and
control units with the objective of using them for the fabrication
of far infrared active cavity radiometers. The sensing unit, first
characterized at \( 300 \)~K using industrial platinium resistance
thermometers, has a noise level of \( \sim 25-30 \)~\( \mu  \)K\( _{eff} \)
for a \( 3 \) hours measuring time and in a \( 1 \)~Hz bandwidth.
Using YBCO superconducting thermometers, the noise level goes down
to \( 2.5 \)~\( \mu  \)K\( _{eff} \), which is strongly limited
by excess 1/f noise in the} \textbf{\textcolor{black}{}}\textcolor{black}{YBCO
film at the superconducting transition. The sample holder used in
the \( 90 \)~K experiments is built with an auxiliary heating resistor,
which enables an easy and accurate identification of the electrothermal
model, even in the closed loop operation. Based on a design previously
published by NIST, we estimate from these experimental results that
the overall noise limitations of radiometers could be lowered by one
order of magnitude.}
\end{abstract}

\newpage

\section{\textcolor{black}{INTRODUCTION}}

\textcolor{black}{The performances of electrical-substitution radiometers
(ESR) depend mainly on low noise temperature control. Such instruments
have been developed for various radiometric measurement purposes,
including, for example, the measurement of the Stefan-Boltzmann constant\( ^{1,2} \),}
and \textcolor{black}{the long-term monitoring of the solar irradiance\( ^{3} \),
and are used as primary standards of optical power at national standards
laboratories\( ^{4,5} \). In these experiments, an optical absorber-receiver
is thermally isolated from a temperature regulated heat sink by a
weak thermal link and is alternately heated by the unknown radiant
power and by the resistive heater (Joule heating). The incident radiant
power entering the aperture of the absorber-receiver can be deduced
from the measurement of the equivalent electrical power needed to
ensure the same temperature rise for the two heating effects. The
resolution of} the \textcolor{black}{ESR was so far mainly limited
by the temperature stability of the receiver. Thus, two temperature
controllers, one for the heat sink and the other for the receiver,
have been proposed to improve the ESR\( ^{6} \). Our work is inscribed
in this framework, and we report in this paper on the progress we
made on the temperature control of the heat sink using a low noise
switched electronics and an optimized analog PID. Furthermore a heat
perturbation method is presented to clearly specify the long time
limitation of the system.}

\textcolor{black}{This paper is arranged as follows. Section 2 describes
the experimental setup and preliminary measurements at room temperature.
The third section relates a set of improvements and measurements at
liquid-nitrogen temperature.} In section 4, \textcolor{black}{the
perspectives for the ESR performance improvements are presented.}

\section{\textcolor{black}{READ-OUT ELECTRONICS AND EXPERIMENTAL SETUP AT
ROOM TEMPERATURE}}

\subsection{\textcolor{black}{Read-out electronics for} \textcolor{black}{\underbar{a}}
\textcolor{black}{standard platinium resistor}}

\begin{figure}[hb]
{\centering \resizebox*{0.8\textwidth}{!}{\includegraphics{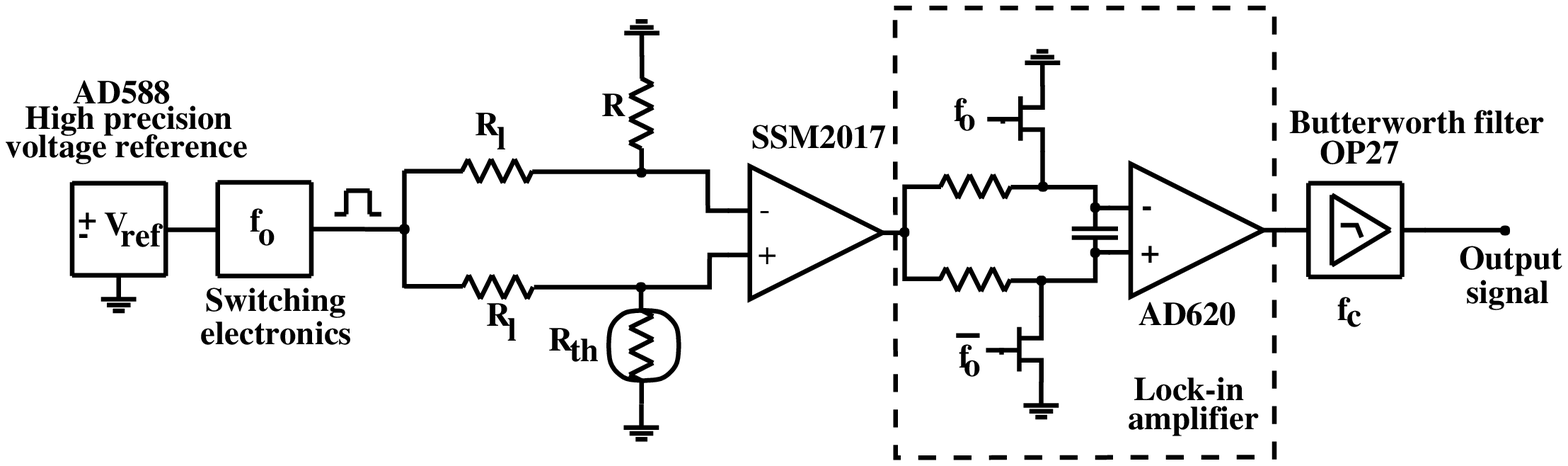}} \par}

\vspace*{50mm}
\caption{\label{Readout}}
\vspace{50mm}
\end{figure}

The fundamental Johnson noise limitation of a metallic thermistor
such as an Industrial Platinium Resistance Thermometer (IPRT), \( 100 \)~\( \Omega  \)
thermistor with \( \alpha =1/R\times dR/dT=3.85\times 10^{-3} \)~K\( ^{-1} \),
leads to an rms value of the noise equivalent temperature (NET) of
\( \sigma =4.2 \)~\( \mu  \)K\( _{eff} \) within a \( 1 \)~Hz
bandwidth (first order) at \( 1 \)~mA bias current, or equivalently,
\( 26 \)~\( \mu  \)K\( _{pp} \) around an ideally stable thermostat
at room temperature. The required resolution for the read-out electronics
is in the \( 10^{-7} \) range (i.e. \( 7^{1/2} \) digits or \( 25 \)
bits resolution). The home made read-out electronics presented here
was designed to be within a factor \( 2 \) of such resolution at
low cost. The read-out electronic system is presented in Fig.~\ref{Readout}. 

An elementary Wheastone DC-bridge (or AC-bridge) is often used in
resistance measurements, and lead resistances can be compensated with
special bridge connections, as in the {}``Mueller bridge{}''\( ^{7} \).
To improve the temperature measurement, different ways have been investigated
in the literature, which decrease the noise equivalent temperature
of the system\( ^{8} \). Our circuit is derived from low-noise electronics
designed for astronomical bolometric measurements. It consists of
a square wave bias current source, a low \textcolor{black}{temperature-coefficient}
resistance, a \textcolor{black}{low-noise} preamplifier stage, a lock-in
amplifier and a \( 2^{nd} \) order Butterworth filter. The periodic
bias current avoids some contamination of the useful signal by low
frequency electrical noise. The square wave bias is used in order
to keep constant the electrical power dissipated in the thermistor,
unlike the sine wave bias\( ^{9} \). Consequently, the temperature
of the thermistor does not oscillate. Indeed, in the sine wave bias
case, a modulated applied power, at \( 2f_{o} \), is dissipated in
the thermistor. Moreover, in order to avoid parasitic harmonic components,
the modulation frequency \( f_{o} \) has to be higher than the bolometer
thermal cutoff frequency \( f_{th} \) (\( f_{o}\gg f_{th} \)).

\textcolor{black}{The high precision voltage reference AD588 was chosen
because of its low output noise, typically \( 6 \)~\( \mu  \)V\( _{pp} \)
over the \( 0.1 \)~Hz~to~\( 10 \)~Hz frequency ranges, of its
long-term stability at ambiant temperature (\( \sim 15 \)~ppm/\( 1000 \)~hours)
and of its low output voltage drift (\( \pm 2 \)~ppm~K\( ^{-1} \)).
It was configured to provide a \( \pm 5V \) reference voltage, noted
\( \pm V_{Ref} \). A switching electronics follows the voltage reference
circuit and then provides the desired square wave voltage at frequency
\( f_{o} \) around a few kHz. As shown in Fig.~\ref{Readout}, the
bias voltage is applied to two high precision resistive dividers,
one of which includes the high sensitivity thermistor \( R_{th} \),
and the other includes a low TCR resistance \( R \) (\( \leq \pm 50 \)~ppm~K\( ^{-1} \),
a few tens of} ohms\textcolor{black}{). It has to be noted that the
electrical power dissipated in the thermometer has to be much lower
than \( 1 \)~mW in order to limit self heating effects\( ^{10} \).
Additionally, the value of the injection resistor, \( R_{l} \), has
to be very large compared to the thermistor resistance \( R_{th} \)
in order to form a current source\( ^{11} \). This injection resistor
was fixed at \( 6 \)~k\( \Omega  \) thus leading to a \( 0.83 \)~mA
current through \( R_{th} \) (\( \sim 100 \)~\( \Omega  \) impedance)
and \( R \).}

\textcolor{black}{An injection resistor was preferred to a capacitance
even if the capacitance is sometimes preferred because it does not
generate Johnson noise\( ^{12} \). In} that \textcolor{black}{case,
the capacitance is fed using a triangular wave voltage, which induces
a square wave current. This triangular wave voltage may be obtained
using an active integrator circuit. Although no Johnson noise takes
place through the capacitance, the system stability is dependant on
the capacitance and the use of an additional amplifier may also increase
the noise level. In order to ensure stability together with the low
noise requirement, we therefore decided to use a square wave voltage
through the low value thermistor and the low TCR reference resistor
\( R \).}

\textcolor{black}{The preamplifier stage is made using an instrumentation
amplifier (SSM2017) followed by a lock-in amplifier. The reference
signals are applied to the lock-in amplifier built around an AD620
and JFET switching connections. Finally, a Butterworth filter (}2nd\textcolor{black}{-order)
shapes the useful output signal of the battery-operated power device.
Using the \( \alpha  \) value of} an IPRT, the \textcolor{black}{voltage-to-temperature
responsivity of the system at the input \( \Re _{i} \) is then \( 320 \)~\( \mu  \)V~K\( ^{-1} \).
The white noise around the frequency \( f_{o} \) is mainly dominated
by the Johnson noise of \( R_{th} \) and \( R \). It can be estimated
as \( e_{n}=\sqrt{2}\sqrt{4k_{B}TR} \) with the \( R \) value chosen
close to \( R_{th} \) at the working temperature. Adding the amplifier
noise, we obtain the equivalent noise source \( e_{n}\simeq 2 \)~nV~Hz\( ^{-1/2} \)
at room temperature. It follows that the ultimate noise equivalent
limit of our system is \( T_{n}=6.3 \)~\( \mu  \)K~Hz\( ^{-1/2} \)
(i.e.} an \textcolor{black}{rms value \( T_{eff}=\sqrt{\frac{\pi }{2}}T_{n}\simeq 8 \)~\( \mu  \)K\( _{eff} \)
in a \( 1 \)~Hz, and a first order bandwith), twice the ideal value
for an IPRT.}

\subsection{\textcolor{black}{Read-out electronic characterization}}

\subsubsection{\textcolor{black}{Noise of the system}}

\textcolor{black}{First, a study of the read-out electronic system
where the thermistor was replaced by an other low TCR resistor has
been carried out. The system exhibited an excess low frequency noise
regim close to a {}``\( 1/f \){}'' one as clearly seen in the comparative
Fig.~\ref{bruit}(c). Then, the deduced noise equivalent temperature
spectrum (NET) would be comprised between \( 6/\sqrt{f} \)~\( \mu  \)K~Hz\( ^{-1/2} \)
and \( 10/\sqrt{f} \)~\( \mu  \)K~Hz\( ^{-1/2} \) if an ideal
IPRT was used at \( 0.83 \)~mA. The associated rms value is in the
range \( 16-30 \)~\( \mu  \)K\( _{eff} \), \( 2 \) times higher
than expected in Sec.~2.1 in the white noise} regime. Deviat\textcolor{black}{ion
from the white noise and the \( 1/f \) law in the read-out electronic
system could be due to the electronic system temperature that slowly
drifts according to the surrounding laboratory temperature. The temperature
compensating coefficients were estimated by measuring the electronic
boards temperature during heating or cooling the whole electronic
systems. Temperature} coefficients \textcolor{black}{of the three
read-out electronic systems were found to be:} \( 15.1\pm 3.6\times 10^{-4} \)~V~K\( ^{-1} \),
\textcolor{black}{\( 5\pm 2\times 10^{-4} \)~V~K\( ^{-1} \) and
\( -7.9\pm 1.4\times 10^{-4} \)~V~K\( ^{-1} \), respectively.
These coefficients are consistent with the observed output drift.
Temperature drifts in the range of \( 40 \)~\( \mu  \)K~s\( ^{-1} \)
at the PC board level were then estimated. They must be avoided or
compensated to minimize the output drifts, which are known to lead
to an excess noise differing from a \( 1/f \) regime, with a higher
\( 1/f^{n} \) power.}

\subsubsection{\textcolor{black}{Room temperature measurement tests}}

\textcolor{black}{The temperature-coefficient} re\textcolor{black}{sistance
(TCR), noted \( \alpha  \), and the dimensionless slope of the thermometer
\( A \), defined as \( A=d\ln R/d\ln T=\alpha T \), are generally
used to compare different thermometers operated at different temperatures.
The Standard Platinium Resistance Thermometer (SPRT) is used to define
the ITS90 (}International Temperature Scale of 1990) \textcolor{black}{between
the triple point of} hydrogen (\textcolor{black}{\( 13.8023 \)~K)
and the freezing point of} silver (\textcolor{black}{\( 1234.94 \)~K)
with an accuracy of \( \pm 2.0\times 10^{-3} \)~K to \( \pm 7.0\times 10^{-3} \)~K\( ^{13} \).
However, the} \textcolor{black}{strain-free} d\textcolor{black}{esign
of a SPRT limits its use in controlled laboratory conditions. In the
industrial platinium resistance thermometer (IPRT), the platinium
wire is encapsulated within a ceramic housing or a thick film sensor
is coated onto a ceramic surface. The protection from the environment
is increased with a metal sheath. A Correge IPRT was used in this
present work. This class A device is characterized by an accuracy
of \( \pm 0.15 \)~K at \( 273 \)~K and an average TCR of \( 3.85\times 10^{-3} \)~K\( ^{-1} \)
(European standard) between \( 273.16 \)~K to \( 373.16 \)~K (according
to DIN CEI 751 norm). This sensing element has a \( 100 \)~\( \Omega  \)
impedance at \( 273.16 \)~K (triple point of water). The room temperature
setup includes three IPRT thermometers radially set out on a good
heat conductor plate (copper), several copper heat shield}s, \textcolor{black}{and
a heater resistance \( R_{H} \). A coolant (oil) is used to minimize
temperature gradients across the system. A correlation study between
two temperature measurements was then possible using this arrangement.
The most important parameters of merit of the thermometers we used
are gathered in Table~}\setcounter{table}{0}\textcolor{black}{\ref{TCRandA}.}

\begin{figure}[hb]
{\centering \resizebox*{0.8\textwidth}{!}{\includegraphics{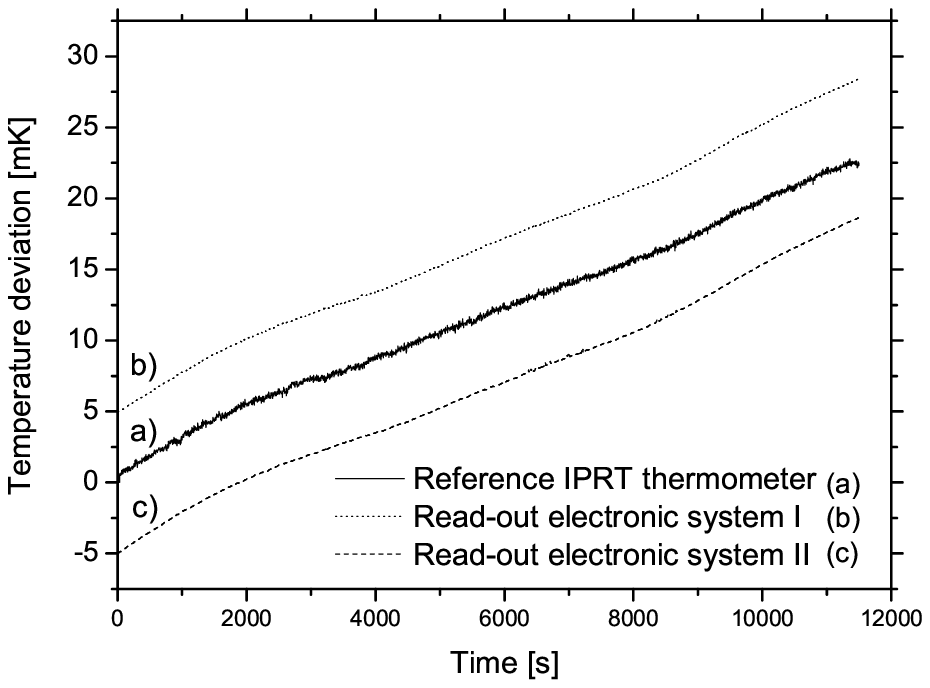}} \par}

\vspace{50mm}
\caption{\label{2methods}}
\vspace{50mm}
\end{figure}

\textcolor{black}{The temperature measurements were made with two
read-out electronic systems, each using an IPRT as thermistor. The
third IPRT was measured using the conventional} four-wire meth\textcolor{black}{od
resistance measurement, performed} with an HP30\textcolor{black}{34A
multimeter (\( 6^{1/2} \)~digits, \( 2 \)~s integration time).
It is hereafter called the reference IPRT thermometer. Responsivity
estimations at the output \( \Re _{o}=\partial V_{out}/\partial T \)
were made using the slow dc drift of the temperature cell, assuming}
\textcolor{black}{\emph{}}\textcolor{black}{that the} \textcolor{black}{\emph{}}\textcolor{black}{reference
IPRT thermometer is well calibrated (see Fig.~\ref{2methods}). Note
the good temperature stability: the temperature drift \( \Delta T \)
was only \( 25 \)~mK for \( 12000 \)~s measurement. Output responsivity
for each read-out electronic system was about \( 10 \)~V~K\( ^{-1} \).}

\begin{figure}[hb]
{\centering \resizebox*{0.8\textwidth}{!}{\includegraphics{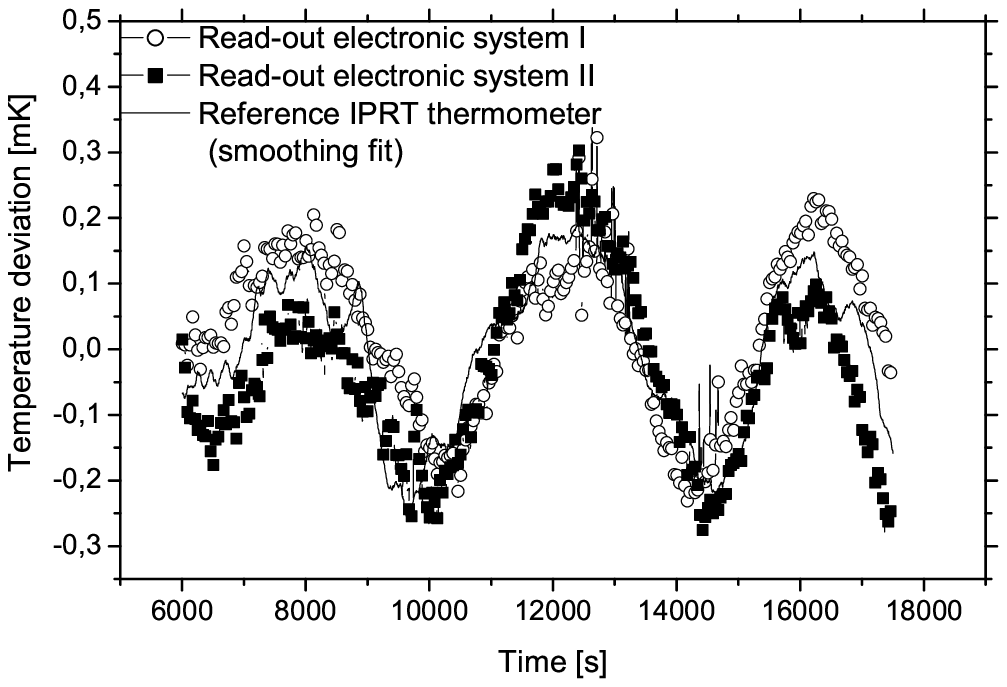}} \par}
\vspace*{50mm}

\caption{\label{correlationIPRT}}
\vspace*{50mm}
\end{figure}

\begin{figure}[hb]
{\centering \resizebox*{0.8\textwidth}{!}{\includegraphics{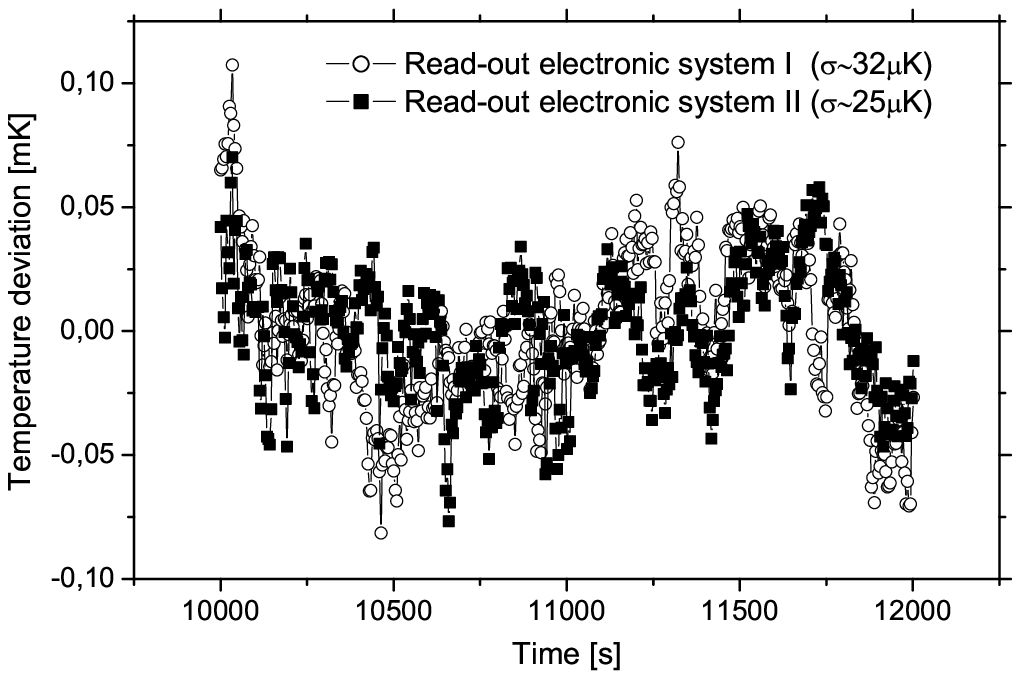}} \par}

\vspace{50mm}
\caption{\label{FluctuationsIPRT}}
\vspace{50mm}
\end{figure}

\textcolor{black}{A fluctuation estimation has been made using a polynomial
fit up to the fourth order to remove the main part of the cell temperature
drift. The result is shown in Fig.~\ref{correlationIPRT}. The fitting
operation explains the quasi-periodic temperature} \textcolor{black}{artifacts
evolutions, but the fact that the three responses are still cohe}rent
between themselves on Fig.~\ref{correlationIPRT} indicates that
the resolution of the thermometers is better than the amplitude of
the \textcolor{black}{artifact} mod\textcolor{black}{ulation, i.e.
\( \sim 200 \)~\( \mu  \)K. To better estimate the resolution of
the high sensitivity read-out, we made a second fit limited to \( 2000 \)~s.
A result is shown Fig.~\ref{FluctuationsIPRT}. This numerical analysis
shows that our read-out system, connected to standard IPRT is compatible
with a temperature resolution of \( 25 \)~\( \mu  \)K\( _{eff} \)
over a \( 2000 \)~s integration time. This value appears to be rather
higher than that what was expected from the theoretical white noise
level as introduced in Sec.~2-1 (\( 8 \)~\( \mu  \)K\( _{eff} \)).
This is due to the excess low frequency noise below \( 1 \)~Hz.
We note that the final estimation of noise is consistent with the
value estimated in subsection 2.2.1 after measurements with a low
TCR resistor. This shows that the excess noise of the IPRT is likely
not seen in this frequency range. Finally we also note that for a
short integration time (\( <1s \)), the} resolution obtained is close
to the theoretical one \textcolor{black}{below \( 10 \)~\( \mu  \)K~Hz\( ^{-1/2} \).
Such values (on short integration time) were also obtained by Dupuis}
\textcolor{black}{\emph{et al.}}\textcolor{black}{\( ^{14} \), but
for \( 100 \)~k\( \Omega  \) thermistors, the intrinsic voltage
noise of which} were \textcolor{black}{\( 31.6 \) times higher than
the \( 100 \)~\( \Omega  \) thermistor used in this study. }

\textcolor{black}{In order to reduce the temperature drifts during
long time measurements, another experimental setup at liquid-nitrogen
temperature was built in association with a home-made optimized PID
controller which will be described in the next section.}

\section{\textcolor{black}{LOW TEMPERATURE SAMPLE HOLDER AND EXPERIMENTS}}

\subsection{\textcolor{black}{Sample holder design}}

\begin{figure}[hb]
{\centering \resizebox*{0.8\textwidth}{!}{\includegraphics{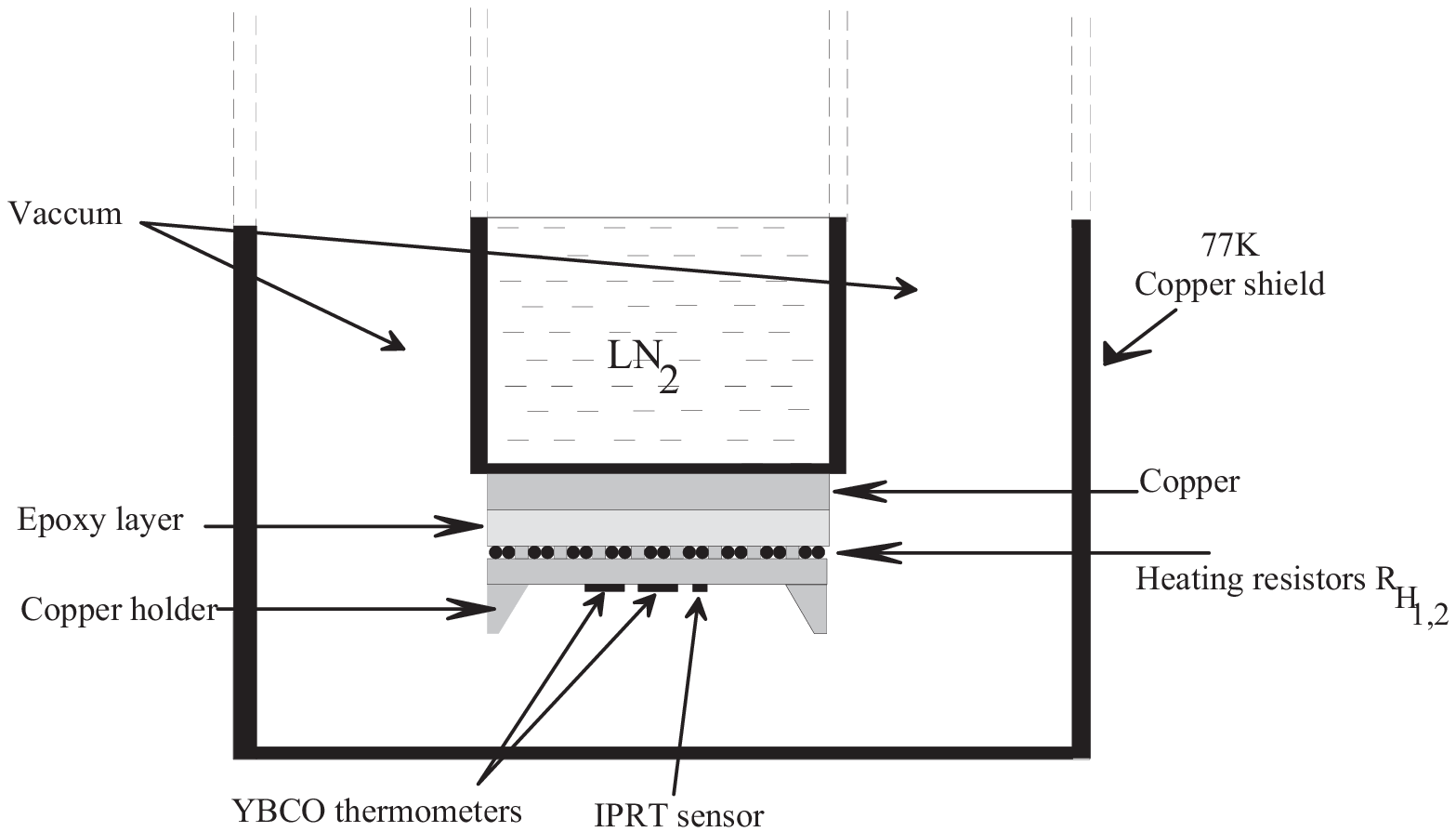}} \par}

\vspace{50mm}
\caption{\label{Setup2}}
\vspace{50mm}
\end{figure}

\textcolor{black}{The sample holder depicted in Fig.~\ref{Setup2}
was designed in order to improve long time temperature measurements
by introducing an optimized temperature control. It is known that
the temperature control of massive systems, even at temperature as
low as \( 77 \)~K, can be made difficult because of the inherent
time delay between the heat production (or heat removing) and the
temperature rise (or temperature decrease) of the sample. It follows
that the overall gain of the servo loop cannot be made arbitrarily
large in order to ensure a sufficient stability (Nyquist criterion).
The present sample holder will serve as the temperature controlled
heat sink of a future active cavity radiometer} (ACR). Its \textcolor{black}{mass
must be much higher than the receiver one, say \( \sim 10^{3} \),
as in the NIST (National Institute of Standards and Technology, USA)
prototype, in order to act as a heat sink\( ^{10} \). We the}n chose
\textcolor{black}{a mass of \( 20 \)~mg for the receiver and \( 20 \)~g
for the sample holder. In order to minimize the heat travel between
heating resistor and the thermometers glued on the copper plate, the
heating resistor, noted \( R_{H_{1}} \), was wound in a spira}l \textcolor{black}{grooved}
at \textcolor{black}{the rear of the plate. The resulting delay \( \tau _{D} \)
is \( 0.4 \)~s. A second heating resistor, noted \( R_{H_{2}} \),
was wound, intertwined and identical to the first one. It constitutes
an easy and reproducible way of applying an accurate heat perturbation
to the system, either in open loop or closed loop configuration. The
use of this auxillary input led to a very convenient identification
of the parameters of the servo-loop models and greatly helped the
full characterization of the system. A large perturbation in the open
loop configuration was then applied in order to determine the thermal
model parameters gathered in Table~\( 2 \). They are those of a
first order thermal circuit with a small delay. Using these values
and the signal characteristics of our read-out electronic system,
we built a SPICE model in order to derive a set of parameters for
the PID controller feedback circuit.}

\subsection{\textcolor{black}{Full system characterization and performances}}

\begin{figure}[hb]
{\centering \resizebox*{0.8\textwidth}{!}{\includegraphics{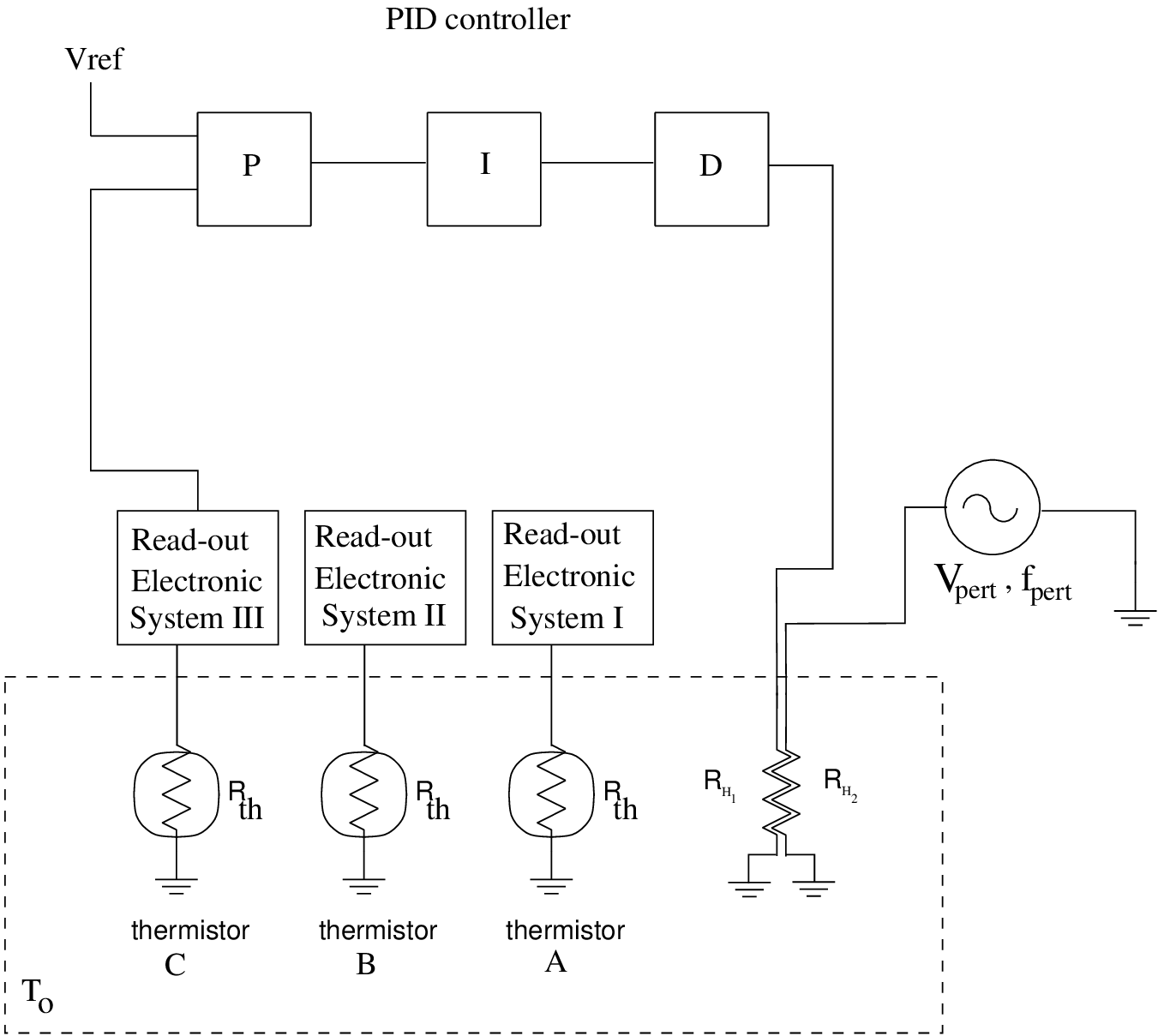}} \par}

\vspace{50mm}
\caption{\label{PID}}
\vspace{50mm}
\end{figure}

\begin{figure}[hb]
{\centering \resizebox*{0.8\textwidth}{!}{\includegraphics{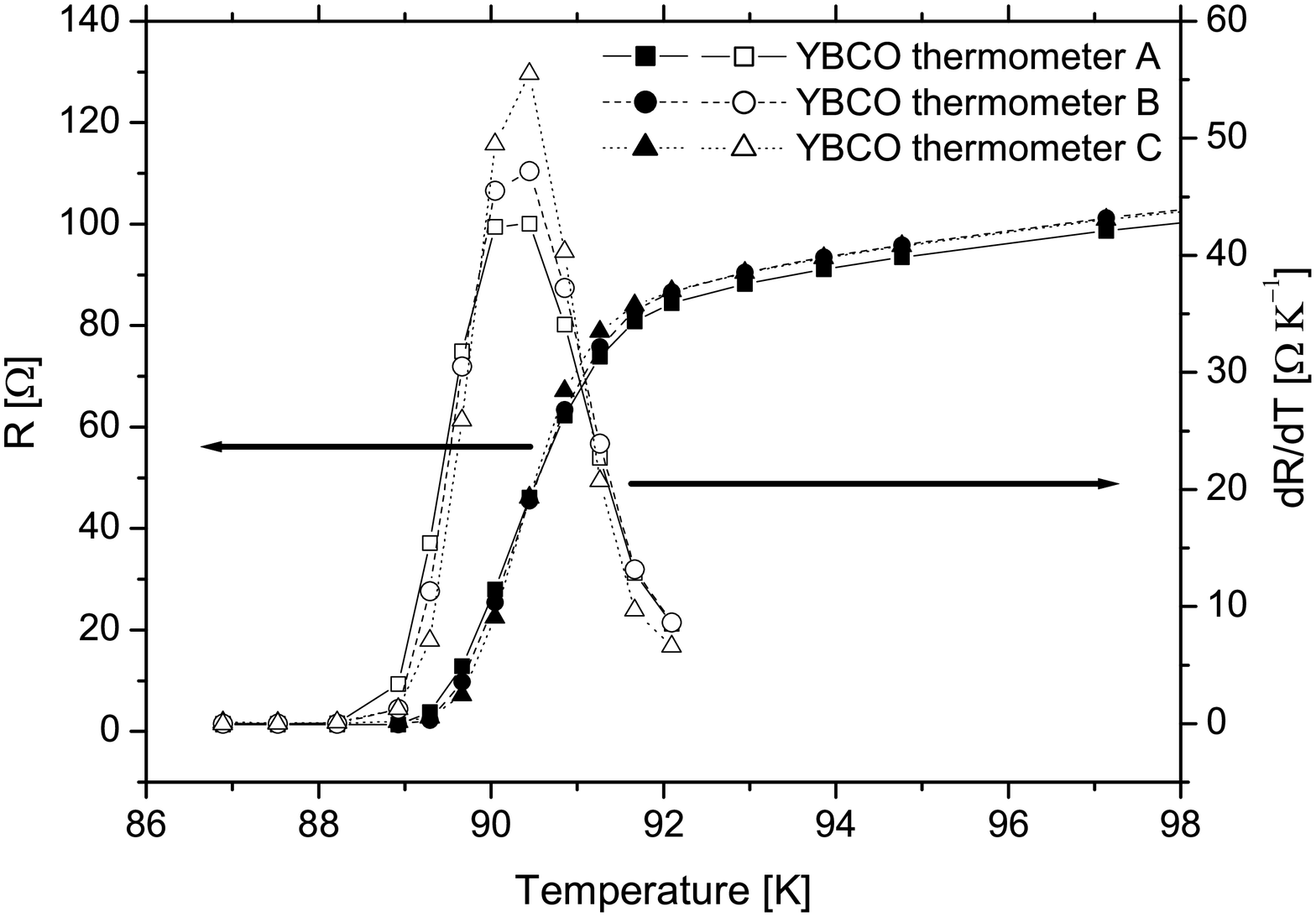}} \par}

\vspace{50mm}
\caption{\label{r(t)}}
\vspace{50mm}
\end{figure}

\begin{figure}[hb]
{\centering \resizebox*{0.8\textwidth}{!}{\includegraphics{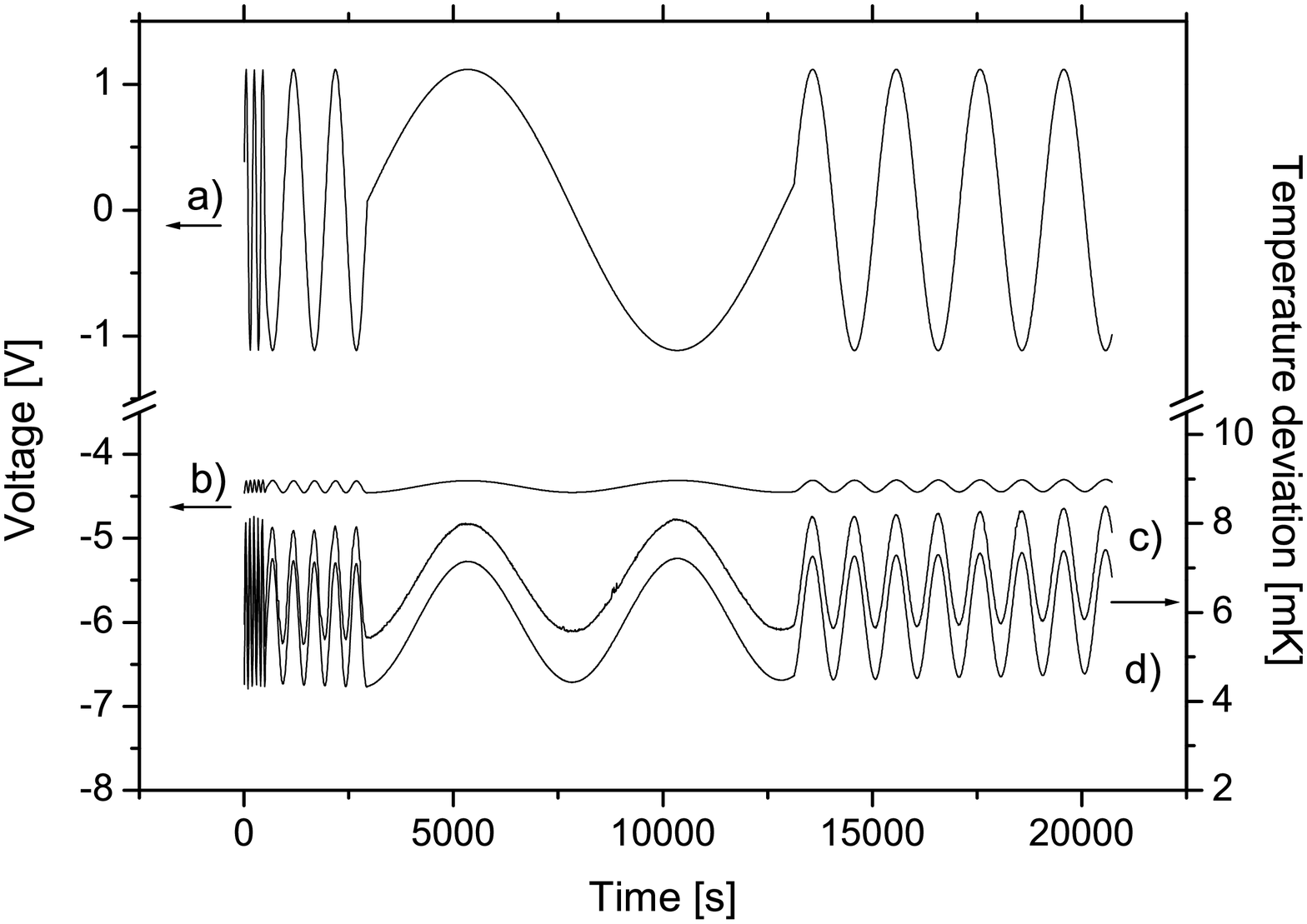}} \par}

\vspace*{50mm}
\caption{\label{perturbation}}
\vspace*{50mm}
\end{figure}

\begin{figure}[hb]
{\centering \resizebox*{0.8\textwidth}{!}{\includegraphics{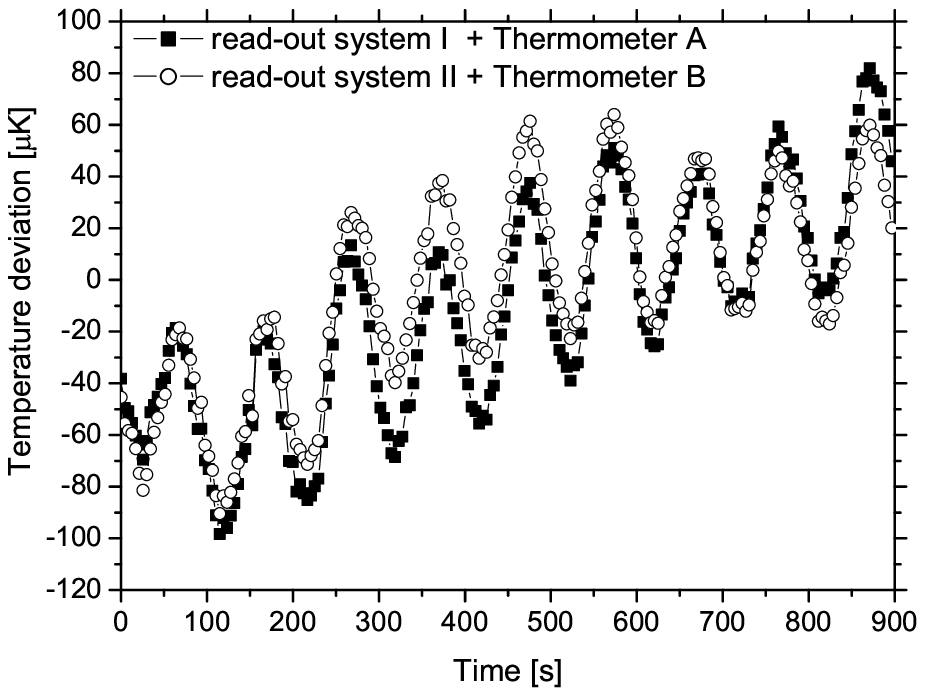}} \par}

\caption{\label{oscillation}}
\vspace{50mm}
\end{figure}

\begin{figure}[hb]
{\centering \includegraphics{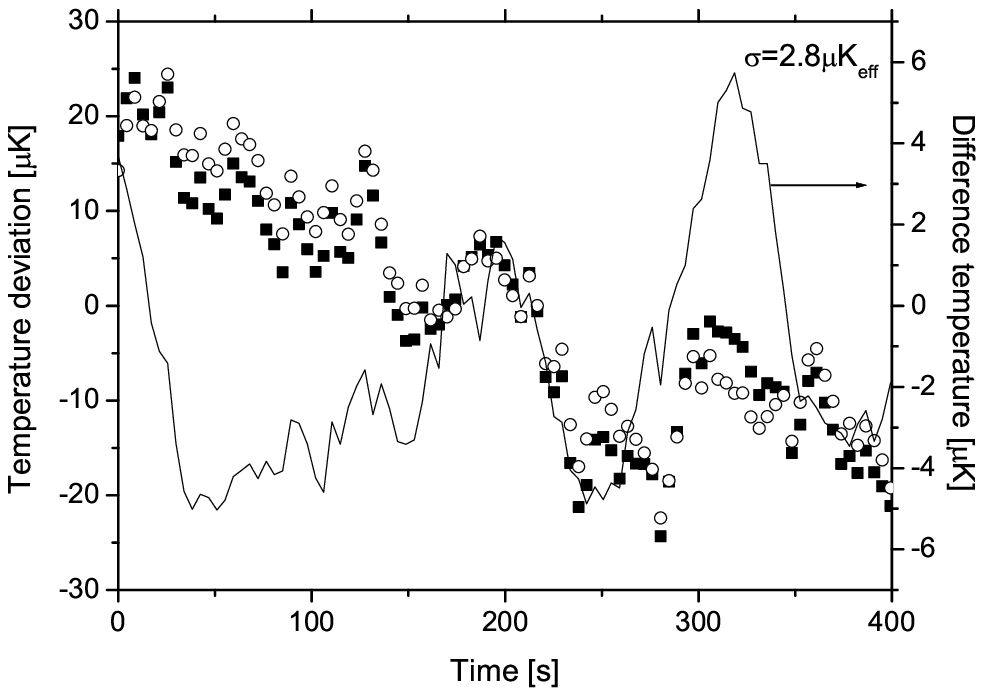} \par}

\vspace*{50mm}
\caption{\label{lowdrift}}
\vspace*{50mm}
\end{figure}

\textcolor{black}{T}he \textcolor{black}{schematic} of the \textcolor{black}{experimental
arrangement used to fully characterize the temperature control of
the sample holder is reported in Fig.~\ref{PID}. We used three high
TCR resistors made using a \( 200 \)~nm thick high quality} \( YBa_{2}Cu_{3}O_{7-\delta } \)
\textcolor{black}{(YBCO) film, patterned in \( 40\times 600 \)~\( \mu  \)m\( ^{2} \)
strips. Their \( R \) versus \( T \) characteristics and their derivatives
are plotted in Fig.~\ref{r(t)}. They exhibit the same critical temperature
and very similar shapes. One of thes}e \textcolor{black}{thermometers}
\textcolor{black}{was used to sense the temperature in the servo loop,
the two others giving two independant observations of the temperature
evolution of the copper plate. Finally, a heat perturbation could
be easily applied by means of the second heating resistor \( R_{H_{2}} \).
Data plotted in Fig.~\ref{perturbation} show examples of this perturbation
method with a large signal applied on \( R_{H_{2}} \). Decreasing
the voltage oscillation amplitude enables the recording of small temperature
oscillations in the two high sensitivity thermometers as shown in
Fig.~\ref{oscillation}.} \textcolor{black}{These} record\textcolor{black}{ings,
performed in longer measuring times, were used to calcula}te \textcolor{black}{the}
noise spectra of our system. The frequency doubling in Fig.~\ref{perturbation}
is of course associated \textcolor{black}{with} the \textcolor{black}{squaring}
of the applied v\textcolor{black}{oltage producing the Joule heating
\( V^{2}/R_{H_{2}} \). From these data, it is easy to extract the
rejection efficiency of the servo loop at these frequencies for the
heat perturbations coupled to the sample holder. We find a value of
\( 400 \) at DC. This means that the stability of the sample holder
temperature is improved by \( \sim 400 \) by closing the loop, and
because the temperature drifts can be \( 70 \)~\( \mu  \)K~s\( ^{-1} \)
in open loop, we are expecting a temperature stability of \( 0.175 \)~\( \mu  \)K~s\( ^{-1} \).
Moreover, the YBCO thermistors have a much higher sensitivity than
the platinium ones, as shown in Table~\( 1 \), leading to a responsivity
\( \Re _{i}\sim 42 \)~mV~K\( ^{-1} \) at \( 0.83 \)~mA bias
current. It follows that the temperature drifts occuring on the PC
board will have a relative effect \( \sim 200 \) times lower than
that obtained using an} IPRT. Then, considering the Johnson noise
of the YBCO thermistors only, \textcolor{black}{the} ideal noise floor
would be about \textcolor{black}{}\( 10 \)~nK\( _{eff} \) in a
\( 1 \)~Hz bandwidth. If the read-out \textcolor{black}{electronics}
is not ideal, the noise floor would be \textcolor{black}{\( 30 \)}~nK\( _{eff} \).
\textcolor{black}{However, the overall performances are limited by
the excess low frequency noise of the YBCO thermistors, the level
of which is generally dependent on microstructural properties. Fig.~\ref{lowdrift}
shows a recording at the output of two high sensitivity thermometers
during a measuring time of \( 400 \)~s. They are clearly correlated,
and show a mean temperature drift of \( 87 \)~nK~s\( ^{-1} \).
This latter value is consistent with the expected one, because the
temperature fluctuations and drifts of the liquid-nitrogen bath are
likely associated} \textcolor{black}{with} the fluctuations and drifts
of the atmospheric pressure with a maximum rate of the order of \( 1 \)~Pa~s\( ^{-1} \)
\textcolor{black}{on} windy days and a conversion \textcolor{black}{coefficient
of \( 83 \)~\( \mu  \)K~Pa\( ^{-1} \) from Clapeyron's law\( ^{15} \).
Finally, using the two simultaneous records of Fig.~\ref{lowdrift},
we also plotted the instantaneous diffe}rence \textcolor{black}{between}
the t\textcolor{black}{wo thermometers, in order to get an estimation
of the actual rms noise of the thermometers. We have got a standard
deviation \( \sigma =2.8 \)~\( \mu  \)K\( _{eff} \) during this
integration time. This value is much higher than the ideal one obtained
using only the Johnson noise assumption. It clearly indicates that
the noise process in our YBCO samples, at the superconducting transition,
is dominated by excess low frequency fluctuations of the resistance.
The associated spectrum has been plotted in Fig.~\ref{bruit}. Note
that the calibration of this spectrum was conveniently made using
a reference signal at \( 5 \)~mHz obtained with a known applied
heat perturbation. Its amplitude was \( 95 \)~mV, producing a temperature
oscillation of the sample holder of \( 30 \)~\( \mu  \)K\( _{pp} \)
in closed loop operation~(see Fig.~\ref{oscillation}). However,
despite this large excess noise, we show in the following section
that these performances are still useful to design an ACR.}

\section{\textcolor{black}{ESR PERSPECTIVES}}

\begin{figure}[hb]
{\centering \resizebox*{0.8\textwidth}{!}{\includegraphics{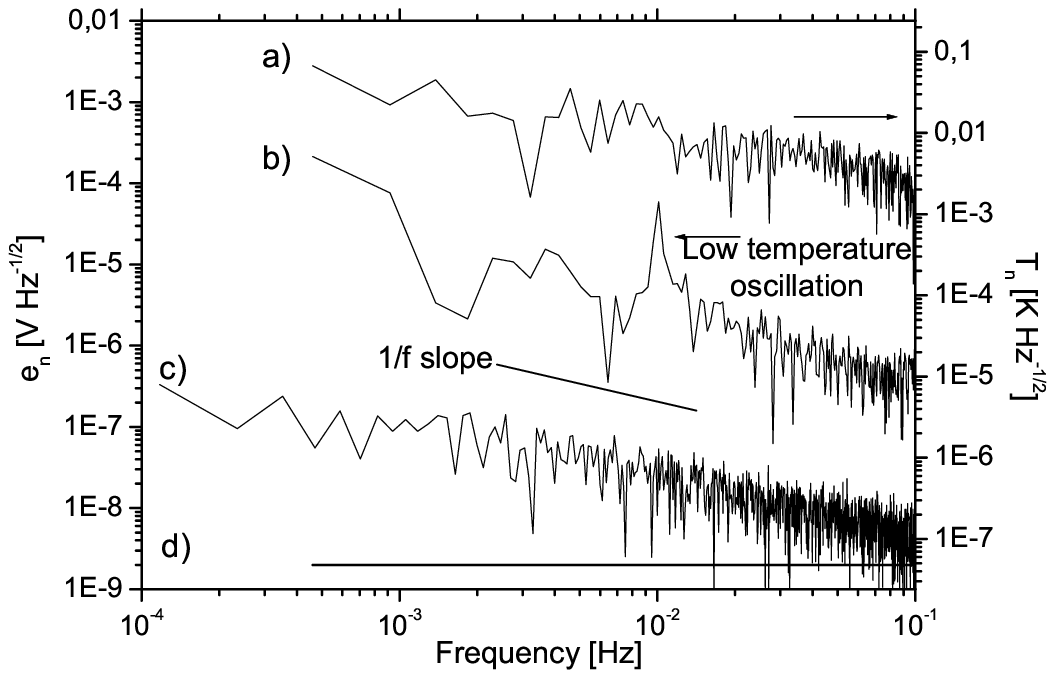}} \par}

\vspace*{50mm}
\caption{\label{bruit}}
\vspace*{50mm}
\end{figure}

\begin{figure}[hb]
{\centering \resizebox*{0.8\textwidth}{!}{\includegraphics{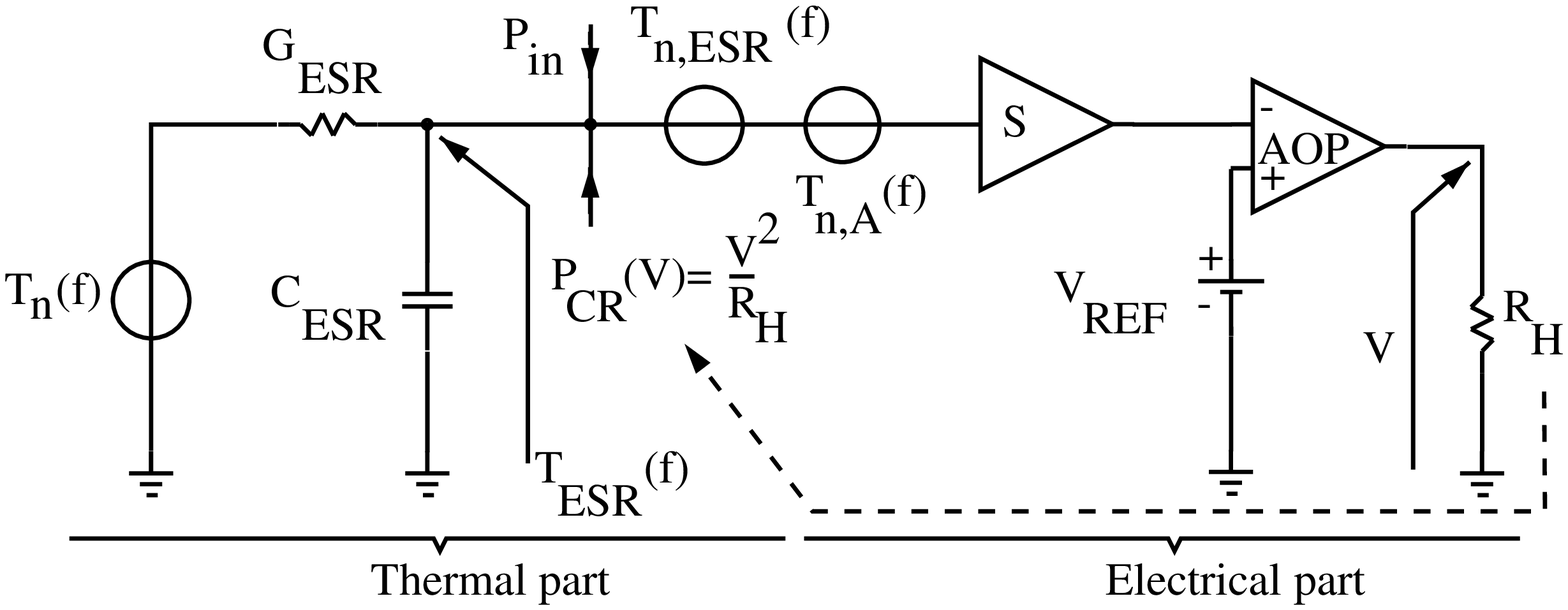}} \par}

\vspace{50mm}
\caption{\label{electrothermal_feedback_circuit}}
\vspace{50mm}
\end{figure}

\textcolor{black}{The achieved low noise temperature control could
be used to regulate the temperature of the heat sink of the ESR. The
NIST prototype described} \textcolor{black}{by Rice~}\textcolor{black}{\emph{et~al.}}\textcolor{black}{\( ^{10} \)}
\textcolor{red}{\underbar{}}\textcolor{black}{will be considered below
as an example. In order to estimate the performances, we will consider
the same conditions as those used in the NIST prototype. The ESR cavity
is supposed to receive radiant flux ranging from the microwatt level
to the milliwatt, coming from an extended-source blackbody at \( T=300 \)~K.
It was shown that the measurement could be done with a contribution
to the standard random uncertainty below \( 20 \)~nW. As explained
in Sec.~3.1, we reduced the mass of the heat sink, which implies
the reduction of that of the receiver down to \( 20 \)~mg. This
condition is thought to be very well fulfilled using a plane membrane
receiver and an integrating sphere above it. To derive our following
noise calculations, we assume that the incident radiation \( P_{in} \),}
t\textcolor{black}{o be converted in electrical power \( P_{CR} \)
in order to keep the receiver at constant temperature, is chopped
at a frequency \( f_{o}=10 \)~Hz. The simple electrothermal} circuit
we used to describe the effects of thermometer \textcolor{black}{noise}
on the receiver is reported in Fig.~\ref{electrothermal_feedback_circuit}
where \( G_{ESR} \) and \( C_{ESR} \) are the ESR thermal conductance
and thermal capacitance, respectively. We introduced the \textcolor{black}{associated
noise sources} to three contributions having roughly the same order
of magnitude:

\textcolor{black}{- \( T_{n}(f), \) which is the spectral density
of the temperature fluctuations of the heat sink. It is shown from
\( 10^{-4} \)~Hz to \( 10^{-1} \)~Hz in Fig.~\ref{bruit}~(b).
Above \( 10^{-1} \)~Hz the servo loop of the heat sink is not suitable
anymore, and we just extrapolated the quasi \( 1/f^{2} \)~\( T^{2}_{n}(f) \)
dependance. The spectral density \( T_{n}(f) \) can be roughly fitted
by \( T_{n}(f)=8.10^{-7}f^{-0,9} \)~K~Hz\( ^{-1/2} \) from Fig.~\ref{bruit},
we estimated its value at \( 10 \)~Hz to be about \( 100 \)~nK~Hz\( ^{-1/2} \).}

\textcolor{black}{- \( T_{n,ESR}(f), \) which is the spectral density
of the YBCO thermometer attached to the ESR. It would act as the error
detector of the servo loop driving the ESR cavity. Because we intend
to use the same material and detecting electronics} \textcolor{black}{as}
that of the heat sink, we adopt the same spectral density \( T_{n,ESR}(f)=T_{n}(f) \),which
should hold up to about \( 100 \)~Hz, before joining the Johnson
level. As a matter of fact, recordings of the \textcolor{black}{thermometer}
output, using a \textcolor{black}{wide opening (\( 300 \)~Hz)} of
the lock-in amplifier unit, led to this \textcolor{black}{spectral
density.}

\textcolor{black}{- \( T_{n,A}(f) \), which is the equivalent noise
associated} \textcolor{black}{with} t\textcolor{black}{he amplifier.
At \( 10 \)~Hz, a reasonable value is given by the ratio of the
voltage noise \( e_{nA}\simeq 1 \)~nV~Hz\( ^{-1/2} \) of the amplifier
to the thermometer responsitivity \( \Re _{i}\simeq 42 \)~mV~K\( ^{-1} \)
at \( 0.83 \)~mA bias.}

\textcolor{black}{Furthermore, in the circuit of Fig.~\ref{electrothermal_feedback_circuit},
\( S \) represents the thermometer sensitivity including gain (in
V~K\( ^{-1} \)). We assume \( S=100 \)~V~K\( ^{-1} \) and AOP
is an operational amplifier, closing the servo loop and feeding the
heating resistor \( R_{H} \) of the ESR cavity}. \textcolor{black}{Assuming
the AOP ideal, the inversing operational amplifier input writes \( V_{-}=V_{REF} \),}
implying \textcolor{black}{the thermometer circuit input to be \( V_{REF}/S \).
Standard circuit analysis leads to: }

\textcolor{black}{\begin{equation}
\label{Tesr}
T_{ESR}(f)=\frac{T_{n}(f)G_{ESR}+P_{in}+P_{CR}}{G_{ESR}+jC_{ESR}\omega }
\end{equation}
}

\textcolor{black}{where the heat} \textcolor{black}{flow} bet\textcolor{black}{ween
the thermometer and the ESR body is neglected. Finally, The {}``Kirchhoff{}''
law gives:}

\textcolor{black}{\begin{equation}
\label{Kirchhoff}
T_{ESR}(f)+T_{n,ESR}(f)+T_{n,A}(f)=\frac{V_{REF}}{S}
\end{equation}
}

\textcolor{black}{The deduced feedback power is written} in t\textcolor{black}{wo
parts:}

\textcolor{black}{\begin{equation}
\label{feedback_power1}
P_{CR}+P_{in}=(G_{ESR}+jC_{ESR}\omega )\frac{V_{REF}}{S}
\end{equation}
}

is the ideal part, if we consider that there is no noise, and:

\textcolor{black}{\begin{equation}
\label{feedback_power2}
P_{CR,n}=-G_{ESR}[(T_{n,ESR}(f)+T_{n,A}(f))(1+j\tau _{ESR}\omega )+T_{n}(f)]
\end{equation}
}

\textcolor{black}{is the noisy feedback power.}

\textcolor{black}{In the bandwidth of the ESR any change in \( P_{in} \)
will induce a change in \( P_{CR} \) in order} to \textcolor{black}{keep
the receiver temperature constant and \( P_{CR} \) accounts for the
output signal}. The second part accounts for the noise on the feedback
power, the minus sign indicating that this part of the feedback power
reacts so as to \textcolor{black}{balance the various noise sources.
Eq.~\ref{feedback_power2} is deduced from a small signal analysis
applied to Eq.~\ref{Kirchhoff}, i.e. by nulling its right member,
and to Eq.~\ref{Tesr}, i.e. with \( P_{in}=0 \). Assuming the noise
sources to be uncorrelated we get the final spectral noise density
of \( P_{CR,n} \):}

\textcolor{black}{\begin{equation}
\label{PCRn}
\left| P_{CR,n}\right| =G_{ESR}\times \sqrt{T_{n}^{2}(f)+\frac{T_{n,A}^{2}(f)+T_{n,ESR}^{2}(f)}{1+j\tau ^{2}_{ESR}\omega ^{2}}}
\end{equation}
We may now use the values attached to the noise sources to estimate
whether or not we fulfill the ESR requirements. For} clarity, \textcolor{black}{we
examine independently the effects of \( T^{2}_{n}(f) \) and \( T^{2}_{n,ESR}(f)+T^{2}_{n,A}(f) \).
The numerical values are calculated at a chopping frequency of \( 10 \)~Hz,
assuming a measuring time of \( 250 \)~s with a first order post
filtering. Before performing these calculations we need to know the
numerical value of the thermal conductance \( G_{ESR} \). The latter
is related to the dynamic range of the ESR, i.e}.\textcolor{black}{,}
\textcolor{black}{the largest input signal \( P_{in,Max} \) that
can be measured. In the chopped mode of the incoming power \( P_{in} \),
the feedback power will oscillate between the on-off states of \( P_{in} \),
leading to the condition of Eq.~\ref{GESR} to be fulfilled by \( G_{ESR} \):}

\textcolor{black}{\begin{equation}
\label{GESR}
G_{ESR}=\frac{S\times P_{in,MAX}}{V_{REF}}
\end{equation}
}

\textcolor{black}{Choosing \( P_{in,MAX}=10 \)~mW, \( V_{REF}=10 \)~V,
\( S=100 \)~V~K\( ^{-1} \) gives \( G_{ESR}=100 \)~mW~K\( ^{-1} \).
Such a thermal conductance value can reasonably} be \textcolor{black}{achieved
at \( 90 \)~K using silicon micromachining techniques\( ^{16,17} \):
for calculations we assumed} \textcolor{black}{a \( 10-30 \)~\( \mu  \)m
thick membrane} \textcolor{black}{of \( 3\times 3 \)~mm\( ^{2} \)
area. It follows that the noise associated with} th\textcolor{black}{at
of the heat sink should be \( G_{ESR}\times T_{n}(10 \)~Hz\( ) \)
of the order of \( 10^{-8} \)~W~Hz\( ^{-1/2} \), i.e}.\textcolor{black}{,
a standard deviation of the order \( 800 \)~pW\( _{eff} \) with
a measuring time of \( 250 \)~s. To estimate the noise associated
with} th\textcolor{black}{at of the receiver thermometer we need to
know the receiver time constant. Because of the small active volume
of the receiver and of the high thermal conductance value, the thermal
time constant would be much shorter than that of the NIST prototype,
about \( 0.1 \)~ms at \( 90 \)~K. We then deduce the order of
magnitude of the associated noise to be close to \( G_{ESR}\times T_{n,ESR}(10Hz)/2\pi \simeq 1.6 \)~nW~Hz\( ^{-1/2} \),
which means an} \textcolor{black}{rms value of \( 130 \)~pW\( _{eff} \)
for \( 250 \)~s observing time. We then conclude that these estimated
values, although estimated using a rough model, are very promising,
within an order of magnitude below the demonstrated values by Rice~}\textcolor{black}{\emph{et~al.}}\( ^{10} \)\textcolor{black}{.
Similar results were presented by Libonate and Foukal}\( ^{18} \)\textcolor{black}{,
reporting a root-mean-square noise level of \( 1.6 \)~nW and a \( 30 \)~s
time measurement for an allowable input-power level of up to \( \sim 2 \)~mW.
These experimental results are of the same order of magnitud}e \textcolor{black}{as}
ours, cal\textcolor{black}{culated for an input-power level of up
to \( 10 \)~mW. No specific optimization in order to reduce the
excess noise in our YBCO films was made, which explains} \textcolor{black}{why}
the noise \textcolor{black}{level} \textcolor{black}{is rather large,
if compared to other works\( ^{17,19} \). As shown by Neff~}\textcolor{black}{\emph{et~al.}}\( ^{17} \),
the use of a thin gold layer \textcolor{black}{onto the YBCO layer
can} l\textcolor{black}{ead to very low noise equivalent temperature
at \( 10 \)~Hz. The use of such levels in our model would have lead
to dramatic reduction of the noise floor of the sample holder, which
means improvements} \textcolor{black}{in} the E\textcolor{black}{SR
performances. We would like to point out that the use of a small membrane
and an integrating sphere does not allow a proper primary radiometer
operation, but the very short time constant would permit the use of
a much higher shopping frequency. This means working conditions in
the white noise domain of the YBCO sensors, implying a possible reduction
of two orders of magnitude of the noise floor level, which appear
still very attractive. Such a system would then have performances
close to the liquid-helium-cooled ESR of Reintsema}\textcolor{black}{\emph{~et~al.}}\textcolor{black}{\( ^{20} \).
Finally, to eliminate the problem of the integrating sphere, we have
done preliminary calculations in order to design a \( 1 \)~cm aperture,
\( 1.5 \)~cm long pyramidal cavity in silicon, wit}h \textcolor{black}{wall
thicknesses} rangi\textcolor{black}{ng between \( 5 \) to \( 50 \)~\( \mu  \)m.
A high conductance value of \( \sim 100 \)~mW~K\( ^{-1} \) is
feasible together with a time constant of the order of \( 0.1 \)~s.
Such a system would then be very attractive because it could include
full-integrated heating resistors and thermometers. Thus correlatio}n
\textcolor{black}{techniques} and \textcolor{black}{a} \textcolor{black}{perturbation}
h\textcolor{black}{eating method could be used as well to improve
model identification and signal processing.}

\section{ACKNOWLEDGMENTS}

This work would not have been possible without the technical contributions
of N. Chéenne and S. Lebargy. We would like to thank Y. Monfort and
M. Lam Chok Sing for many stimulating conversations. 

\newpage

\section*{References}

\( ^{1} \)T.J. Quinn and J.E. Martin, Philos. Trans. R. Soc. London
\textbf{A316}, 85 (1985)

\( ^{2} \)A. C. Parr, NIST Tech. Note \textbf{1421} (1996)

\( ^{3} \)J.P. Rice, S.R. Lorentz and T.M. Jung, \textcolor{red}{\emph{\underbar{}}}\textcolor{black}{Proc.
10th Conf. on Atmospheric Radiation}\emph{,} Amer. Meteor. Soc., 85
(1999)

\( ^{4} \)N. P. Fox, Metrologia \textbf{37}, 507 (2000)

\( ^{5} \)C. C. Hoyt and P. V. Foukal, Metrologia \textbf{28}, 163
(1991)

\( ^{6} \)J.P. Rice and Z.M. Zhang, NIST Tech. Note \textbf{1414}
(1996)

\( ^{7} \)\textcolor{black}{H. Sostmann, Isotech Journal of Thermometry
(1989)}

\( ^{8} \)P.R.N. Childs, J.R. Greenwood and C.A. Long, Rev. Sci.
Instrum. \textbf{71}, 2959 (2000)

\( ^{9} \)A.E. Lange, S.E. Church, P.D. Mauskopf, V. Hristov, J.J.
Bock, H.M. DelCastillo, J. Beeman, P.A.R Ade and M.J. Griffin, Proc.
30th ESLAB Symp., 105 (1996)

\( ^{10} \)J.P. Rice, S.R. Lorentz, R.U. Datla, L.R. Vale, D.A. Rudman,
M. Lam Chok Sing and D. Robbes, Metrologia \textbf{35}, 289 (1998)

\( ^{11} \)J.C. Mather, Appl. Opt., 1125 (1982)

\( ^{12} \)S. Gaertner, A. Benoît, J.M. Lamarre, M. Giard, J.L. Bret,
J.P. Chabaud, F.X. Désert, J.P. Faure, G. Jegoudez, J. Landé, J. Leblanc,
J.P. Lepeltier, J. Narbonne, M. Piat, R. Pons, G. Serra and G. Simiand,
A.\&A. Suppl. Ser. \textbf{126}, 151 (1997)

\( ^{13} \)H. Preston-Thomas, Metrologia \textbf{27}, 3 (1990)

\( ^{14} \)P. Dupuis and C. Eugène, Proc. 6th IMEKO TC-4 Symposium,
309 (1993)

\( ^{15} \)M. Lam Chok Sing, E. Lesquey, C. Dolabdjian and D. Robbes,
IEEE Trans. Appl. Supercond. \textbf{7}, 3087 (1997)

\( ^{16} \)L. Méchin, J.C. Villégier and D. Bloyet, J. Appl. Phys.
\textbf{81}, 7039 (1997)

\( ^{17} \)H. Neff, I. A. Khrebtov, A. D. Tkachenko, E. SteinbeiB,
W. Michalke, O. K. Semchinova, T. Heidenblut and J. Laukemper, Thin
Solid Films \textbf{324}, 230 (1998)

\( ^{18} \)S. Libonate and P. Foukal, Metrologia \textbf{37}, 369
(2000)

\( ^{19} \)D. G. McDonald, R. J. Phelan, L. R. Vale, R. H. Ono and
D. A. Rudman, IEEE Trans. Appl. Supercond. \textbf{9}, 4471 (1999)

\( ^{20} \)C. D. Reintsema, J. A. Koch and E. N. Grossman, Rev. Sci.
Instrum. \textbf{69}, 152 (1997)

\newpage

\section*{Tables}

\begin{table}[hb]

\caption{\setcounter{table}{1}\label{TCRandA}Parameters of merit of the studied
thermometers: \protect\( T_{o}\protect \) is the operating temperature,
\protect\( \Delta T\protect \) is the measurement temperature range,
\protect\( \alpha \protect \) is the temperature coefficient resistance
(TCR) and \protect\( A\protect \) is the dimensionless slope of the
thermometer.}

\begin{longtable}{lcccc}
&
&
&
&
\\
\hline
\hline 
studied thermometers&
\( T_{o} \)&
\( \Delta T \)&
\( \alpha  \)&
\( A \)\\
&
{[}K{]}&
{[}K{]}&
{[}K\( ^{-1} \){]}&
{[}dimensionless{]}\\
\hline 
IPRT sensor&
\( 273.16 \)&
\( 13.8023 \) to \( 1234.94 \)&
\( 0.00385 \)&
\( \sim 1 \)\\
\( YBa_{2}Cu_{3}O_{7-\delta } \) &
&
&
&
\\
(typical values)&
around \( 90 \)&
few K&
\( 1 \) to \( 5 \)&
\( 90 \) to \( 450 \)\\
\textcolor{black}{(}present work\textcolor{black}{)}&
\( 90.25 \)&
\( 2-3 \)&
\( 1.7 \)&
\( 150 \)\\
\hline
\hline 
&
&
&
&
\\
\end{longtable}
\end{table}

\begin{table}[hb]

\caption{\setcounter{table}{2}\label{thermal_parameters}Thermal parameters
and heating resistors for the liquid-nitrogen temperature measurement
setup.}

\begin{longtable}{lcc}
&
&
\\
\hline
\hline 
parameter&
symbol&
value\\
\hline
mean thermal time constant of the sample holder&
\( \tau _{Th} \)&
\( 100 \)~s\\
thermal conductance between heat sink and copper plate&
\( G_{Th} \)&
\( 77 \)~mW~K\( ^{-1} \)\\
delay between applied electrical heating and temperature plate&
\( \tau _{D} \)&
\( 0.4 \)~s\\
main heating resistor&
\( R_{H_{1}} \)&
\( 30 \)~\( \Omega  \)\\
auxiliary heating resistor&
\( R_{H_{2}} \)&
\( 30 \)~\( \Omega  \)\\
\hline
\hline 
&
&
\\
\end{longtable}
\vspace{30mm}
\end{table}

\newpage

\section*{Figures captions}

\textbf{FIG. 1.} Read-out electronic system: a square wave voltage
is applied to two voltage dividers, one includes the thermistor \( R_{th} \),
the other includes a low TCR resistance. A true differential amplifier
(SSM2017) followed by a lock-in amplifier and a Butterworth filter
(2nd-order) shapes the useful output signal of the battery-operated
power system.

\textbf{FIG. 2.} Simultaneous temperature evolution of two read-out
electronic systems (curves b and c), each using an IPRT sensor as
thermistor: the slow temperature drift enables the evaluation of the
responsivity \( \Re  \) . A reference IPRT thermometer (curve a)
is used to calibrate the temperature deviation.

\textbf{FIG. 3.} Temperature measurements after the substraction of
the main temperature drift, showing the correlation between the three
thermometer \textcolor{black}{r}esponses: read-out electronic systems,
each using an IPRT as thermistor (symbols), and a smoothing fit of
reference IPRT thermometer (line)

\textbf{FIG. 4.} Fluctuation temperature during \( 2000 \)~s after
a linear fit, measured with two read-out electronic systems each using
an IPRT as thermistor. A standard deviation \( \sigma  \) of \( 25 \)~\( \mu  \)K
is demonstrated (for system II). A \( 100 \)~\( \mu  \)K maximal
temperature difference is observed between the two thermometer read-outs.

\textbf{FIG. 5.} The liquid-nitrogen temperature measurement setup:
a reference IPRT thermometer and YBCO thin film thermometers are glued
on the copper sample holder. Two heating resistors \( R_{H_{1,2}}=30 \)~\( \Omega  \)
were made using constantan wires and distributed evenly across the
sample holder cross-section.

\textbf{FIG. 6.} Schematic of the home-made PID controller: the three
boxes named P, I and D depict the functions proportional, integral
and derivative, respectively. \textcolor{black}{The temperature measurements}
were made with three read-out electronic systems, each using an YBCO
thermometer: one was used for regulation purpose (thermistor C) and
two for correlation investigations (thermistors A and B). Two heating
resistors \( R_{H_{1,2}}=30 \)~\( \Omega  \) were made using constantan
wire.

\textbf{FIG. 7.} Resistances \( R \) (closed symbols) and their derivative
\( dR/dT \) (opened symbols) as a function of temperature \( T \)
for three high resolution YBCO thermometers at \( 1 \)~mA bias. 

\textbf{FIG. 8.} A sine wave voltage, called perturbation heating,
is applied to the heating resistor \( R_{H_{1}} \)(curve a). The
PID output reacts as shown in curve b (note the frequency doubling).
The resulting temperature deviation is measured by the two read-out
electronic systems, each using an YBCO thermometer (curves c and d,
right axis), which show very similar responses.

\textbf{FIG. 9.} Thermometric responses of the read-out electronic
systems associate\textcolor{black}{d} with \textcolor{black}{a low}
temperature oscillation provided by a \( 5 \)~mHz perturbation heating
(\( \sim 95 \)~mV amplitude) applied to the sample holder.

\textbf{FIG. 10}. Temperature measurement without perturbation signals
using two read-out electronic systems, each using an YBCO thermometer
: the two temperature deviation are plotted (left axis). The instantaneous
difference (right axis) gives an estimation of the actual rms noise
of the thermometers. For a \( 400 \)~s measurement, a standard deviation
equivalent of \textcolor{black}{\( 2.8 \)~\( \mu  \)K\( _{eff} \)}
is obtained.

\textbf{FIG. 11.} Input spectral noise density \( e_{n} \) and noise
equivalent temperature (NET) \( T_{n} \) for a read-out electronic
system without thermistor (curve c) and with an YBCO film (curve b)
as well as the theoretical white noise value (curve d) are plotted.
The NET of the \emph{}reference IPRT thermometer is reported for comparison
(curve a on right axis only).

\textbf{FIG. 12.} Simple electrothermal circuit used to describe the
effects of thermomet\textcolor{black}{er} noise \textcolor{black}{on}
the receiver: \( G_{ESR} \) et \( C_{ESR} \) are the ESR thermal
conductance and therma\textcolor{black}{l} capacitance\textcolor{black}{,
r}espectively, \( P_{in} \) and \( P_{CR} \) are the incident radiation
and the electrical power, respectively. \( T_{n}(f) \) is the spectral
density of the temperature fluctuations of the heat sink. \( T_{n,ESR}(f) \)
is the spectral density of the YBCO thermometer attached to the ESR.
\( T_{n,A}(f) \) is the noise equivalent temperature associat\textcolor{black}{ed}
with \textcolor{black}{the} amplifier.

\newpage

\end{document}